\documentclass[a4paper]{jpconf}
\usepackage{graphicx}
\usepackage{psfrag}
\begin{document}
\title{Zero--temperature study of vacancies in solid $^4$He}

\author{M Rossi, E Vitali, D E Galli and L Reatto}

\address{Dipartimento di Fisica, Universit\`a degli Studi di Milano, 
         via Celoria 16, 20133 Milano, Italy}

\begin{abstract}
We are investigating the properties of vacancies in solid $^4$He with the exact zero temperature
SPIGS method.
Our aim is to study the possibility of phase--separation between vacancies and the perfect 
crystal.
We find a significant correlation among vacancies both in two and in three dimensional solid 
$^4$He.
In 3D we have considered up to 5 vacancies in a hcp crystal at a density $\rho = 0.0293$\AA$^{-3}$,
close to the melting, and in 2D we have considered up to 6 vacancies in a triangular crystal at 
$\rho = 0.0765$\AA$^{-2}$.
In all the considered cases the correlation among vacancies seems to display an exponential decay
that would suggest the presence of a bound state.
The systematically sublinear dependence on the number of vacancies found for the activation
energy strengthens the indication of an attractive interaction, even if we have direct evidence
that vacancies do not form a single compact cluster.
\end{abstract}

Vacancies in a quantum solid like solid Helium have attracted a great interest because of their 
unique properties~\cite{Simm}: extensive theoretical and experimental work on solid Helium has 
indicated that vacancies have an high probability of tunneling from site to site due to the large 
zero--point motion of the atoms.
This distinguishes helium from other solids in which the vacancies are localized at low 
temperature.
Despite of this large effort, our understanding of vacancies in solid Helium is not 
complete~\cite{Simm,Good}.
Moreover, much of the interest in vacancies in solid $^4$He resulted from theoretical 
speculation~\cite{Andr} that the vacancy concentration could be finite even at zero temperature
(zero--point vacancies) and that this could result in a Bose--Einstein condensate (BEC) in the 
solid~\cite{Andr,Ches}.
In that case, solid $^4$He would be a supersolid and it could display some superfluid properties
such as non classical rotational inertia (NCRI) \cite{Legg}.
Zero--point vacancies have revealed to be elusive to the experimental observation~\cite{Simm,Sim2}. 
On the contrary, NCRI effects have been recently observed in solid $^4$He~\cite{Chan}, giving a 
renewed interest to the supersolid state of matter~\cite{Bali}.
Even for this topic, our understanding is far from being complete~\cite{Bali,Prok}.

Interaction among vacancies in solid $^4$He have been recently studied by means of an elastic 
theory~\cite{Maha}. 
Here we systematically investigate the properties of vacancies in solid $^4$He by means of a full 
microscopic technique such as the Shadow Path Integral Ground State (SPIGS)~\cite{spigs}.
The SPIGS method allows to recover ``exact'' expectation values on the true ground state of $^4$He
systems, and it is particularly indicated for non homogeneous and defected systems.
We have considered vacancies both in three dimensional (3D) and in two dimensional (2D) solid $^4$He.
The study of vacancy properties in 3D solid $^4$He at $T=0$K is relevant for the possible supersolid
phase.
In fact, if vacancies form a tight--bound state, as recently suggested by finite temperature PIMC
simulations~\cite{Boni}, they would separate breaking the mechanism that would induce BEC~\cite{Gall}.
The 2D system represents a good reference model for solid $^4$He adsorbed on planar 
substrates, such as graphite, which are under experimental investigation, and it is also relevant
for its conceptual counterpart: the Abrikosov lattice of flux lines in type-II 
superconductors~\cite{Frey}.

Dealing with low temperature properties, $^4$He atoms are described as structureless zero--spin
bosons, interacting through a realistic two--body potential, that we assume to be the HFDHE2 
Aziz potential~\cite{Aziz}.
The Path Integral Ground State (PIGS)~\cite{pigs} method recovers more and more accurate 
approximations of the true ground state wave function by successive short time projections in 
imaginary time of a trial wave function.
The true ground state expectation values will be reached only in the limit of infinite imaginary 
time; in practice, being the convergence exponential~\cite{Bon2}, after a suitable number of small 
time projections the PIGS values differ from to the true ones within the statistical errors and
further projections affect no more the computed values.
In this sense, the results provided by PIGS are exact.
Moreover we have recently shown that the PIGS method is unaffected by any variational bias due 
to the choice of trial wave function~\cite{Vita}.

When a shadow wave function (SWF)~\cite{Viti,Moro} is employed as trial wave function, the SPIGS 
method is recovered~\cite{spigs}.
The use of SWF is justified both by the fact that it gives the best available variational 
description of solid and liquid $^4$He~\cite{Moro} and by the fact that no equilibrium positions 
for the solid phase are requested allowing to easily describe also disordered systems~\cite{spigs}.
We have employed the pair--product approximation for the imaginary time projector~\cite{Cepe}
and the imaginary time step $\delta\tau=1/40$K$^{-1}$ has been chosen in order to ensure a good 
accuracy and a reasonable computational effort~\cite{Vita}. 

\begin{figure}[b]
 \includegraphics[width=9cm]{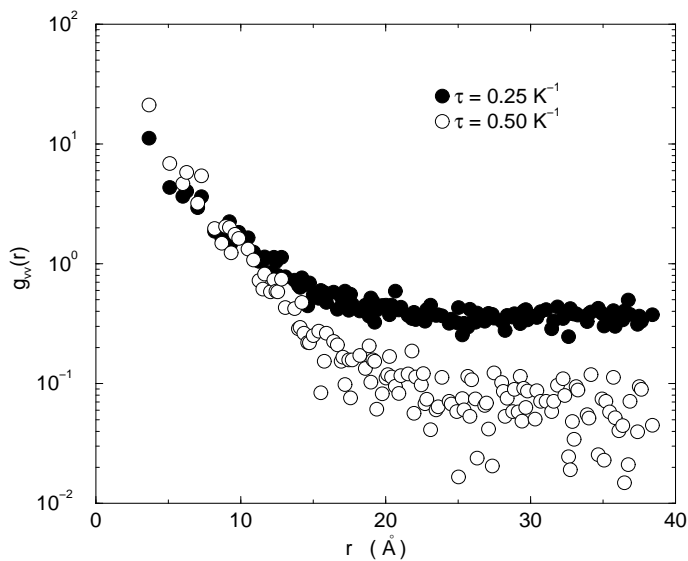}
 \hspace{0.9cm}
 \begin{minipage}[b]{6cm}
  \caption{\label{f:gvv3D} Vacancy--vacancy correlation function $g_{vv}(r)$ computed in a hcp
           crystal with $M=720$ lattice positions and $n=3$ vacancies at the density
           $\rho = 0.0293$\AA$^{-3}$ for two different imaginary projection time $\tau$.
           $g_{vv}(r)$ is normalized with the correlation function of 3 vacancies randomly
           distributed on the same lattice.}
 \end{minipage}
\end{figure}

In order to calculate the activation energies of vacancies, we perform a number of independent 
simulations: a simulation for the perfect crystal and a simulation of the crystal with $n$ 
vacancies for different $n$ values.
In presence of vacancies, after removing $n$ particles, we rescale the dimensions of the simulation
box to reset the system to the original density.
This is performed to circumvent the need of correcting the energy due to density change caused
by the inclusion of vacancies.
Thus the activation energy of $n$ vacancies in a crystal with $M$ lattice sites is 
$\Delta E_n=[e(M-n)-e(M)](M-n)$ where $e(N)$ is the energy per particle for the system containing
$N$ particles.
Starting from an initial configuration corresponding to a perfect lattice in which $n$
particles have been removed, we find that, in all the considered cases, the crystalline state
is stable and the periodicity is such that the number of maxima in the local density remains
equal to $M$.
This means that we have a crystal with $n$ mobile vacancies.
When dealing with the three dimensional solid $^4$He we have considered a box with periodic
boundary conditions (pbc) built to fit an hcp crystal at $\rho = 0.0293$\AA$^{-3}$ with $M=180$ 
lattice positions.
We have computed the activation energy for up to 5 vacancies.
We find $\Delta E_1=15.4\pm0.8$K, $\Delta E_2=30.6\pm0.8$K, $\Delta E_3=43.9\pm0.8$,
$\Delta E_4=53.7\pm0.8$K and $\Delta E_5=63.4\pm0.7$K.
We can see that the dependence of $\Delta E_n$ from $n$ is systematically sublinear, suggesting the
existence of some kind of attractive interaction among vacancies.
A similar result is obtained also in the two dimensional case.
We have considered a box with pbc housing a triangular crystal at $\rho = 0.0765$\AA$^{-2}$ 
with $M=240$ lattice positions and a number of vacancies from 1 to 6.
The obtained activation energies are $\Delta E_1=7.0\pm0.3$K, $\Delta E_2=12.2\pm0.5$K, 
$\Delta E_3=15.5\pm0.3$, $\Delta E_4=19.1\pm0.3$K and $\Delta E_6=25.1\pm0.4$K.
Even in this case, the dependence of $\Delta E_n$ on $n$ is systematically sublinear.

In order to infer whether this interaction is strong enough to give rise to a bound state and to a
phase--separation we have computed also a vacancy--vacancy correlation function $g_{vv}(r)$
by histogramming the relative distances of the vacancies during the Monte Carlo sampling.
The determination of the vector positions of the vacancies in a crystalline configurations is 
far from being trivial due to the large zero point motion, to high vacancy mobility and because 
in our algorithm the center of mass is not fixed. 
The first step of our analysis is, given a configuration explored by the Metropolis sampling, 
to find the crystal lattice $\{\vec R^{(0)}_1\}$ which best fits the positions of the real particles
$\{\vec r_j\}$.
This is achieved by optimizing the position of the center of mass of the reference crystal lattice 
in order to maximize the Gaussian local density 
$\sum_{j=1}^N\sum_{i=1}^M\exp[-\alpha(\vec r_j-\vec R^{(0)}_i)^2]$,
where $\alpha$ is a parameter suitably chosen in order to have the best efficiency.
The position of a vacancy is then recovered by coarse graining procedure, which allocates in subsequent
steps each particle to an exclusive lattice position by considering larger and larger distances, up to
a maximum distance $2a$ ($a$ is the lattice parameter).
If it turns out impossible to allocate all the particles within a distance of $2a$ from a 
reference lattice position, the configuration is discarded.
In order to ensure the convergence of $g_{vv}(r)$ we have considered two different starting
configurations: one with the $n$ vacancies at random lattice positions in the crystal, and
one with the $n$ vacancies in a compact cluster configuration.

\begin{figure}[b]
 \includegraphics[width=9cm]{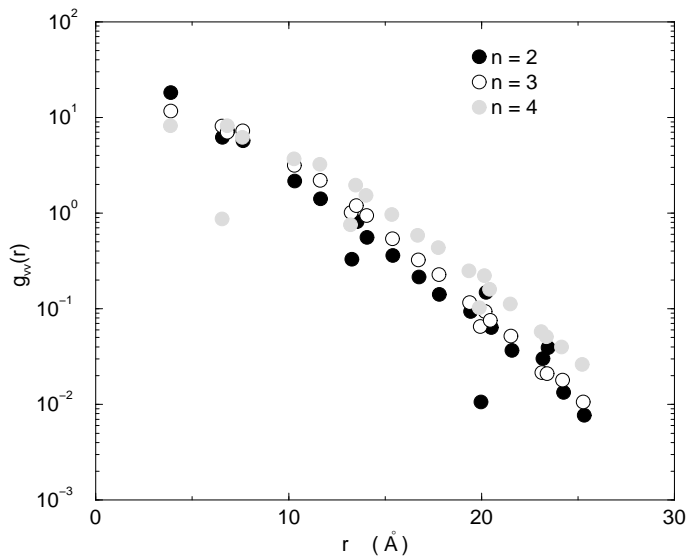}
 \hspace{0.9cm}
 \begin{minipage}[b]{6cm}
  \caption{\label{f:gvv2D} Vacancy--vacancy correlation function $g_{vv}(r)$ computed in a
           triangular crystal with $M=240$ lattice positions and $n=2$, 3 and 4 vacancies at
           the density $\rho = 0.0765$\AA$^{-2}$.
           The total projection imaginary time is $\tau=0.775$K$^{-1}$.
           Correlation functions computed with larger $\tau$, not shown here, do not show
           significant departures from the plotted ones.
           $g_{vv}(r)$ is normalized with the correlation function of $n$ vacancies randomly
           distributed on the same lattice.}
 \end{minipage}
\end{figure}

Our results for 3 vacancies in a 3D hcp crystal with $M=720$ lattice positions at
$\rho = 0.0293$\AA$^{-3}$ are plotted in Fig.~\ref{f:gvv3D} for two different $\tau$ values.
The large values of $g_{vv}(r)$ found in the short distance range enforce the suggestion of a 
strong attractive interaction among vacancies, moreover $g_{vv}(r)$ shows an exponential decay 
for distance up to about 15\AA.
Despite of the substantial depletion at large distances, vacancies are able to explore the whole 
available distance range, as shown by the large distance plateau of $g_{vv}(r)$.
Our present data seem to be not compatible with the phase--separation picture.
However, whereas up to 10\AA~$g_{vv}(r)$ seems at convergence with respect to the $\tau$ value, 
this is not true for the tails where $g_{vv}(r)$ still depends on $\tau$ (see Fig.~\ref{f:gvv3D}).
Computations with larger $\tau$ are underway. 

Similar results are found also in the two dimensional case.
Our results for 2, 3 and 4 vacancies with a total projection time $\tau=0.775$K$^{-1}$
(larger than in the 3D case) are reported in Fig.~\ref{f:gvv2D}.
Contrarily to the 3D results, no large distance plateaus are found in $g_{vv}(r)$,
suggesting a bound state.
When $n=2$, $g_{vv}(r)$ turns out to be well described by an exponential decay with correlation 
length $\lambda\simeq3$\AA.
For larger $n$, $g_{vv}(r)$ shows a more complex behavior, and it seems possible to recognize two 
regimes characterized by different $\lambda$ values.
For example, when $n=4$, $\lambda\simeq14.3$\AA~for $r<12$\AA~and $\lambda\simeq2.97$\AA~for 
$r>12$\AA.
This large $r$ behavior could be significantly affected by the pbc, simulations with larger boxes 
are underway.
It should be noticed that the same $g_{vv}(r)$ is recovered both when the simulations starts with 
the vacancies placed in a compact cluster configuration and randomly placed in the crystalline 
lattice.
Thus, even if initially placed in a compact cluster configuration vacancies explore the whole 
available distance range: this is a direct evidence that vacancies do not form a compact cluster.
As already pointed out, our algorithm discards some configurations where it fails to allocate the
particles to the reference lattice sites within the maximum distance of $2a$.
By looking at these configurations in the 2D case for the larger $n$ considered values, we 
can see that sometime the system rearranges the point defects into couples of dislocations, which 
in 2D are still point defects.
The appearance of such topological excitations is observed also in classical 2D solids~\cite{Ling}.

To conclude, we have presented a study of vacancies in solid $^4$He in 3D and 2D crystals.
We have computed the activation energy $\Delta E_n$ for different numbers of vacancies $n$ and we
have found a systematic sublinear dependence of $\Delta E_n$ on $n$, which gives the indication of an
attractive interaction between point defects.
This is quite similar to what is found for classical solids~\cite{Lech}.
Following the positions of the vacancies during the Monte Carlo sampling we were able to compute
a vacancy-vacancy correlation function $g_{vv}(r)$.
For all the considered cases $g_{vv}(r)$ seems to display an exponential decay which suggests that
vacancies form a bound-state.
But, at the same time, vacancies are always able to explore the whole available distance range
hence do not form a compact cluster.
Further investigations are needed in order to have more quantitative details on vacancy--vacancy
interactions.

\end{document}